# Molecular Spin Qu*d*its for Quantum Algorithms

Eufemio Moreno-Pineda,[a] Clément Godfrin,[b,c] Franck Balestro,[b,c,d] Wolfgang Wernsdorfer*[a,b,c] and Mario Ruben*[a,e]

Presently, one of the most ambitious technological goals is the development of devices working under the laws of quantum mechanics. One prominent target is the quantum computer, which would allow the processing of information at quantum level for purposes not achievable with even the most powerful computer resources. The large-scale implementation of quantum information would be a game changer for current technology, because it would allow unprecedented parallelised computation and secure encryption based on the principles of quantum superposition and entanglement. Currently, there are several physical platforms racing to achieve the level of performance required for the quantum hardware to step into the realm of practical quantum information applications. Several materials have been proposed to fulfil this task, ranging from quantum dots, Bose-Einstein condensates, spin impurities, superconducting circuits, molecules, amongst others. Magnetic molecules are among the list of promising building blocks, due to (*i*) their intrinsic monodispersity, (*ii*) discrete energy levels (*iii*) the possibility of chemical quantum state engineering, and (*iv*) their multilevel characteristics, leading to the so called Qu*d*its ($d > 2$), amongst others. Herein we review how a molecular multilevel nuclear spin qubit (or qu*d*it, where $d = 4$), known as $TbPc_2$, gathers all the necessary requirements to perform as a molecular hardware platform with a first generation of molecular devices enabling even quantum algorithm operations.

## Introduction

The account of the non-independent description of quantum states, i.e. entanglement and superposition, dates back to the time of Einstein[1] and Schrödinger.[2] This phenomenon led the renowned physicist Richard Feynman[3] amongst others[4] to propose the exploitation of entanglement and/or superposition of states to perform certain computational tasks not achievable with classical computers. However, the challenge of such a proposal was the development of both hardware and software components, i.e. building blocks and algorithms, compatible to perform processing of information at the quantum level.

On the side of quantum algorithms, the early developments of the quantum information processing (QIP) field were slow until as late as 1994 when Peter Shor reported a remarkable quantum polynomial algorithm to factorise integers.[5] This was quickly followed by Grover's quantum algorithm[6], which proposes to use quantum mechanics to achieve quadratic speedup in search queries of unsorted data bases. Later on, Lloyd validated Feynman's proposition where a quantum computer (QC) could simulate intractable quantum systems.[7] Undoubtedly, these advantages led to one conclusion: a QC and QIP can outperform classical processing schemes, computers or algorithms in certain tasks that even extremely powerful computer clusters would not be able to achieved.

On the hardware side, the thorough governance by the quantum mechanical laws ultimately require new classes of materials as building blocks. In this context, during the last decade several materials have been proposed as so-called quantum bits (qubits) ranging from defect in solids[8-10], quantum dots[11,12], photons[13,14], impurities in solids[15-17], superconducting systems[18,19], trapped ions[20,21], magnetic[22-25] and non-magnetic[26,27] molecules amongst others. From this library of prospect qubits, molecular quantum magnets, also termed single molecule magnets (SMMs), are very promising systems, due to their appealing magnetic characteristics and facile manipulation *via* chemical means[25,28] which allows the modification of the ligand field of the spin carriers as well as the interaction with other units.[29-31] In one of the several schemes, the long-lived nuclear spin states of these molecules, embedded in the respective devices, are used to encode and to process quantum information, while their electronic states are used to couple and to read-out the quantum information. Amongst the many magnetic molecules available, the single molecule magnet $TbPc_2$ complex has shown by far all the necessary properties to be embedded as active unit into a series of quantum mechanical devices such as molecular spin valve[32], resonator[33], and transistor.[34,35]

Moreover, magnetic molecules possess unique advantages such as tailored chemical control over their surroundings, chemical engineering of the involved quantum states, and defect-free production leading to an infinite number of atomically precise copies. Herein we discuss the quantum characteristics of a prototype molecular qubit, namely $TbPc_2$. It will be shown that these properties make magnetic molecules plausible candidates to perform quantum algorithms; one of the most ambitious targets of the field. It is the scope of this review to also highlight the importance of the nuclear spins embedded in SMMs, which, when coupled to the electronic properties of SMMs, could well represent very versatile nuclear spin qubits and qudits.

## Elementary unit for Quantum Information Processing

### Qubits

The basic unit of information in classical computation is known as the bit comprising two states, 0 and 1. During a computational task, specific operations are carried out by the conversion of sequences of bits, leading to specific operations. In a classical computer, each set of bits has a physical meaning, which can operate as e.g. ON/OFF, or TRUE/FALSE depending on the value of the state.

Analogously, the basic unit of a quantum computer is the quantum bit or qubit, which can perform the exact same operations using two well-defined states i.e. $|1\rangle$ or $|0\rangle$. These states are commonly represented on the Bloch sphere as an arrow pointing to the north pole for the $|0\rangle$ state or south pole

a. Institute of Nanotechnology (INT), Karlsruhe Institute of Technology (KIT), Hermann-von-Helmholtz-Platz 1, D-76344 Eggenstein-Leopoldshafen, Germany. E-mail: mario.ruben@kit.edu; wolfgang.wernsdorfer@kit.edu
b. Université Grenoble Alpes, Institut Néel, F-38042 Grenoble, France
c. CNRS, Institut Néel, F-38042 Grenoble, France
d. Institut Universitaire de France, 103 boulevard Saint-Michel, 75005 Paris, France
e. Institut de Physique et Chimie des Matériaux de Strasbourg (IPCMS), CNRS-Université de Strasbourg, 23 rue du Loess, BP 43, F-67034 Strasbourg Cedex 2, France for $|1\rangle$ state (Fig. 1a). Interestingly, due to the quantum nature of the qubit, the superposition of the $|1\rangle$ or $|0\rangle$ states can be generated with $|\psi\rangle = a_0|0\rangle + a_1|1\rangle$ pointing in any direction of the sphere where the squares of $a_0$ and $a_1$ are the amplitude of the probability following $|a_0|^2 + |a_1|^2 = 1$ (Fig. 1a). These new states have no classical analogues and represent non-orthogonal configurations with $2^N$ states, where $N$ is the number of qubits. This is one of the properties, which gives QC the potential power to perform immensely large and complex operations. For example, it has been estimated that if we were to write in classical numbers all the superposition of states obtained from 300 qubits, we would end up with more numbers than atoms in the entire universe.[36] Extension of the qubits concept, when more than two levels are involved within a given system, leads to the so-called qu*d*its, where the *d* stands for the multilevel characteristics.[37-39] Additionally, qu*d*its offer $d^N$ orthogonal states, allowing parallelisation in a single unit with lower error rates *cf*. qubits counterparts.

*Classical and Quantum Gates*

In classical computation operations are carried out through logical gates, implementing Boolean functions, which yield a deterministic single output. An example of such Boolean operation is the classical NOT gate, which receives an input bit value and sheds a single bit output. Contrary to classical gates, quantum gates allow for the superposition of the states involved during the operation.[40] An example of such quantum gate is the single qubit Hadamard gate, which allows the superposition of states. That is, if $n$ qubits are prepared at a given initial state and then a Hadamard gate is applied to each of the qubits, a total of $n$-qubits superposition is obtained, representing all possible combinations of the $n$ qubits from 0 to $2n - 1$, i.e. $H|0\rangle \otimes H|0\rangle \otimes \cdots \otimes H|0\rangle = \frac{1}{\sqrt{2^n}}\Sigma_{j=0}^{2^n-1}|j\rangle$. The final superposition of states contains all possible solutions of a given problem, acting as shortcuts accelerating the computation process (Fig. 1b). This gate is one of the most important in quantum computing, since it offers the possibility of create exponentially many states, using just polynomial operations.

**Requirements for a Qubit: The DiVincenzo Criteria**

As could be inferred, any two-level system could in principle act as a physical platform for a qubit. In practice, however, several criteria must be met by a material before it can be considered a plausible qubit, e.g. coherence has to be prepared within a given life time, amongst other requisites. These requirements were described in a seminal work by DiVincenzo.[41] Accordingly, any two level system must gather five criteria in order to be considered a viable qubit candidate. These can be summarised as follows:

(*i*). *A scalable well-defined system*: the considered systems must exhibit two well-defined levels. Electron and nuclear spins with $s$ = ½ or $I$ = ½ respectively have been proposed given that these represent *true* two-level systems. Furthermore, systems comprising an effective doublet ground state, ($S_{eff}$ = ½), can also constitute valid candidates as long as the ground doublet state is well isolated from excited states. Superconducting quantum bits (SC), ultracold atoms and several molecular magnetic materials fall within this category. Multilevel state systems qubits, also termed qu*d*its, have likewise been proposed as mimics for intrinsically interconnected qubits to perform quantum algorithms exploiting their multilevel characteristics (*vide infra*).

(*ii*). *Long coherence times*: during information processing, coding is accomplished *via* the coherent state of the qubit, which in order to be manipulated has to live long enough so that the operation is completed. However, qubits are exceptionally susceptible to interactions with surroundings enhancing decoherence, which perturbs the states during computation, leading to information lost. Therefore, the amount of physical interactions must be limited to gain coherence times longer than the time needed to carry out logic gate operations. A figure of merit of $\tau_d/\tau_g$ lower than $10^{-4}$ has been proposed (where $\tau_d$ stands for decoherence time and $\tau_g$ stands for gate time operation).

(*iii*) *Initialisation*: the qubit must be brought into a well-defined state before starting the manipulation. Depending on the nature of the material acting as qubit several initialization procedures can be used, e.g. temperature, light, electromagnetic fields, etc.

(*iv*). *Universal quantum gates*: selectively addressing the qubits to perform algorithms *via* entanglement and superposition of states is a requirement. Thus, two-qubit gates (qugates) are sufficient to perform quantum algorithms. Interestingly, a Hadamard gate allows the gate operation in a single multilevel qubit or qu*d*it, without requiring entanglement. For example, the realisation of the Grover's algorithm is solely based on superposition of states[42], therefore this algorithm can be achieved by creating a superposition of many states through a Hadamard gate in a single qu*d*it (*vide infra*).

(*v*). *Read out*: In partial conflict with requirement (*ii*) stating the isolation of the qubit (or qugate) from the environment, it is essential that after the successful quantum operation the outcome could be read out. That is, to obtain the result after an algorithm has been carried out, by a very *controlled* application of some physical stimuli, e.g. electric fields, light, magnetic fields, etc.

DiVincenzo later added two other requirements[41], described in an article dedicated to quantum communication: (*vi*) interconversion between stationary and dynamic "flying" (e.g. photonic) qubits and (*vii*) exact transmission of "flying qubits". These two extra requirements are important for the successful transmission of information by using entanglement of photons and are of relevance for non-local qubits.[14] Due to this non-local characteristic of the requirements, we do not describe them in detail in this review.

**Quantum Algorithms**

The exploitation of quantum mechanical laws in computing and information processing has been in the mind of scientists since 70's, with many theoretical proposals of the outstanding capabilities that just QIP would afford.[3,43] Despite this, however, experimental investigation towards QIP only boosted around mid-nineties when Peter Shor reported a quantum algorithm able to factorise integers with a quadratic speed up, leading to the first practical application of QIP[44] and by the Grover's algorithm for data queries in unsorted data bases[6]. In the following section a brief description of the most important quantum algorithm will be revised.

*Shor's factorisation algorithm*

In general, an algorithm can be described as a set of instructions that can be used to solve a specific problem. Depending on the complexity of the set of instructions they could be very fast to incredibly slow. The factorisation of integer numbers is an example of a slow algorithm, where the best classical algorithm takes a time $t = \exp(O(\log N)^{1/3}(\log \log N)^{2/3}))$ to calculate the factor of an *N* integer. Shor, in his pioneer work, showed that through quantum mechanical resources only a time $O(\log N)^3$ was needed to solve the problem[44], thus demonstrating the pitfalls of the public-key Rivest-Shamir-Adleman (RSA) cryptographic system.[45] In other words, with a quantum computer, working under quantum mechanical laws, any system operating within the RSA security scheme would be vulnerable to attacks. The solution to this problem using classical means is very difficult. Kleinjung and co-workers demonstrated that a factorisation of a 768-bit number was achieved after a period of two years employing hundreds of computers and a total of $10^{20}$ operations.[46] Conversely, predictions of a factorisation of a 2,000-bit number employing a quantum computer would require solely over a day, using $\sim 3\times 10^{11}$ gates, highlighting the power that quantum resources would offer to demanding tasks.[47]

Experimentally, the realisation of Shor's factorisation algorithm has been achieved for $N = 15^{27,48}$ and $21.^{49}$ The first account of the experimental demonstration of Shor's algorithm came about seven years after Shor's report by Vandersypen *et al.* employing seven nuclear spins of a (2,3-$^{13}$C)hexafluorobutadiene molecule.[27] The nuclear spins act as qubits, whilst the manipulation to conduct the algorithm was achieved *via* nuclear magnetic resonance pulse sequences; thus, successfully factorising $N = 15$, with solution 3 and 5. Later it was shown that factorisation of $N = 15^{48}$ and $21^{49}$ can also be carried out employing photons as qubits.

*Grover's search and optimisation algorithm*

Soon after the proposal of the Shor's algorithm for factorisation of integers, in 1996 Lov Grover devised an algorithm which yields quadratic speedup for a search query in a random database. The problem hypothesized by Grover consisted in the data search in an unsorted database with a number $N = 2^n$ items, where the best classical algorithm would require a total of $O(N)$ operations. Due to the coherence of states, Grover's algorithm would however require $O(\sqrt{N})$ permutations, a quadratic speed up, to find the correct solution.

Remarkably, Loss and Leuenberger in 2001 proposed SMMs to act as qubits to perform the Grover algorithm, making use of the multilevel resources of these entities, without requiring the common prerequisite of entanglement.[42]

Experimentally, proofs of Grover's algorithm have been achieved employing two-qubits molecules such as partially deuterated cytosine molecule where two $^1$H atoms act as qubits.[50] Three-qubits Grover's algorithm has also been achieved employing an isotopically $^{13}$C-labelled CHFBr$_2$ molecule, where the nuclear spins of $^1$H, $^{19}$F and $^{13}$C are initialised, manipulated and read-out through NMR sequences[51] and in a single nitrogen vacancy in diamond possessing an electron spin ($S = 1$) coupled to the nuclear spin of $^{14}$N through hyperfine interaction ($I = 1$) manipulated *via* multi frequency pulses.[52]

Besides Shor and Grover algorithm, which exemplify the usefulness and power of QIP, several other algorithms have been described, such as the Deutsch–Jozsa (D-J) quantum algorithm[26], able to distinguish between a balanced and unbalanced function. Alongside these examples, probably the most important characteristic of quantum information processing is the true simulation of quantum systems, not achievable with current state-of-the-art computers.[3,7] Example of the usefulness of quantum systems was realized employing trapped ions to simulate the dynamic of spin systems.[53] The adventurous reader interested in quantum algorithm subject is refereed to more specialised literature dedicated to the topic.[54]

## Qubit's Materials

Since DiVincenzo established the fundamental requirements for an operational qubit, a wide range of physical platforms have been tested such as: ultracold atoms[20,21], photons[14], superconductors[18,19], defect in solids (such as nitrogen vacancies (NV) in diamond[8-10]) impurities in materials[15,16,55] and molecular[29,34,35,56,57] systems are amongst the most studied materials (Fig. 2).

All these materials, under the right conditions, are two-level systems, which can give coherence and can be initialised through a variety of external stimuli. Nonetheless, the preparation of non-molecular based systems demands very often advanced lithographic processes and complex micro-electronic setups for the isolation of the effective two levels and to achieve low decoherence rates. Moreover, the isolated nature of some of these qubits, in particular those based on nuclear spins, renders the entanglement to other qubits a challenging task, therefore imposing an important problem for the realisation of quantum gates.

Molecular materials also form part of the prospect qubits, owing to their tailored chemical control and monodispersity. These systems have been studied since the early developments of quantum computing field, and currently offer advantages compared or even better than non-molecular systems (*vide infra*). For example, the physical implementation of molecular systems in hybrid devices would speed up the information processing, since the interactions with other qubits (*i.e.* spatial distribution, important for the realisation of quantum gates) can be systematically controlled *via* chemical means; a major drawback observed defects in solids systems such as NV in diamond. Additionally, after Loss and Leuenberger report, molecular materials became even more appealing, due to their discrete multilevel energy level characteristics. The foremost advantage of the qu*d*its ($d > 2$) over their two-level qu*b*its ($b = 2$) counterparts is due to their ability to realise processing of information at quantum level in a single unit with diminished error rates. Additionally, entanglement and superposition could be achieved in qu*d*its in large dimensions with smaller clusters of processing units compared to conventional qubits.

Generally speaking, one of the main sources of decoherence is related to the interaction with the surrounding. To quantify the coherence of a qubit two parameters have been thoroughly studied in molecular qubits, $T_1$ and $T_2$, where $T_1$ embodies spin-lattice relaxation time, and $T_2$ is the spin-spin relaxation, representing the coherence lifetime of the qubit. Intriguingly, although seemingly contradictory, interaction with other qubits is necessary for the successful implementation of logic gates, thus information processing. The successful realisation of a quantum gates employing two level systems requires entanglement: quantum gates are applied on the resultant "large" Hilbert space arising from the entanglement between the two units. In this sense, molecular magnets have shown promising characteristic: long coherence times comparable to alternative physical platforms[35,56,58-60]; moreover, it has been shown by Winpenny and Affronte that controlled communications can be achieved between the two-qubits located on the same molecule[26].

In contrast, when employing multilevel systems (qu*d*its), each level of the qu*d*its represents a dimension (spanning the Hilbert space), thus entanglement is not necessary for the realisation of a Hadamard gate. In this case, the multilevel character of the qu*d*its can be exploited to perform quantum gates in a single qu*d*its ($d > 2$) unit, through a Hadamard gate, creating a superposition of the many level states.[27,61,62] In addition, multilevel qu*d*its are more advantageous than two-level qubits counterparts, due to their ability to parallel quantum information processing in a single unit, decreasing the error rates.[38,63-66] Furthermore, entanglement and superposition could be achieved in qu*d*its in large dimensions with smaller clusters of processing units.

*Electronic Spin Molecular Qubits*

Molecular electron spin qubits[38,63-68] have been proposed since they are examples of $S = \frac{1}{2}$ systems, where electronic spins can be easily addressed through application of moderate temperature and magnetic fields. The most widely proposed mean of manipulation for electron-spin qubits is via pulse Electron Paramagnetic Resonance (EPR). A very well-known family of electron spin qubit candidates are the {$Cr_7Ni$} antiferromagnetic wheels, extensively studied by Winpenny's group, and subject of innumerable studies, owing to the antiferromagnetic coupling between the $Cr^{3+}$ and $Ni^{2+}$, which leads to a well-defined spin $S = \frac{1}{2}$ ground state, isolated from excited states.[69] The {$Cr_7Ni$} family has demonstrated long $T_1$ and $T_2$ times determined through pulse EPR. Remarkably, $T_1$ and $T_2$ can be enhanced by chemical variation of the environment of the {$Cr_7Ni$} wheel[58], while it remains very robust when linked to one[29,31,70] or more[71] units, leading to their proposal for logic gates.

Detailed study of the effect on $T_1$ and $T_2$ has revealed that the main source of decoherence at molecular level can be ascribed to the nuclear spin bath where the qubit is immersed: the nuclear spins of atoms which form part of the chelating ligands and surrounding solvent molecules. The rational synthesis that molecular systems offer have allowed the detection of impressive $T_1$ and $T_2$ times even at room temperature and in bulk crystals[59,60]. For example, Bader et al. reported coherence times up to 1 μs at room temperature in a chemically engineered copper molecule by systematic elimination of nuclear spins.[59] Additionally, Freedman's[56,72-74] and Sessoli's[60,75,76] groups have turned their attention towards vanadyl based complexes, leading equally to the observation of long coherence times at even room temperature. Fig. 3 shows some examples of spin-spin relaxation time ($T_2$) studies of several electron spin qubits prospect. As observed, the values are comparable to convenctional materials such as NV centres.[77]

Besides 3d-metal containing systems, lanthanides have also been proposed as electronic qubits[30,57,78] owing to the Kramers doublet characteristics with inherent magnetic anisotropy and separation of the ground doublet from the first excited states. Based on this property, Aromí and co-workers proposed an asymmetric lanthanide dimer as two qubit molecular candidate, where the small interaction between the $Ce^{3+}$ and $Er^{3+}$ lanthanide ions could in principle be addressed through manipulation of the resonance frequencies or fields, leading to the proposal of a CNOT gate.[30]

Unfortunately, as earlier mentioned, electron spin units are extremely susceptible to interactions with the spin bath, i.e. electron spins are strongly contingent on factors such ligand and solvent characteristics. Furthermore, most of these impressive results have been obtained in crystals on large ensembles, which would make the initialisation step very challenging, due to inhomogeneity effects. Likewise, the highly vulnerable character of electron spins threatens the ultimate goal of devices constructions, where their deposition between leads and/or on surfaces is deemed to introduce further interactions.[79]

*Nuclear Spin Molecular Qubits*

The inherent shielding from the environment yielding extremely long coherence times and very low error rates, make nuclear spins promising as qubits.[80] Alas, due to the small magnetic moments and close nature, interactions are very weak, making problematic their integration to circuits, read-out and manipulation. However, it has been shown that despite these challenging characteristics, the read-out and manipulation of nuclear spins can be achieved.[34,35]

The molecular qubit subject of the following sections of this review comprises a central $Tb^{3+}$ octa-coordinated ion sandwiched between two parallel aligned phthalocyanines, exhibiting a square-antiprismatic ($D_{4d}$) coordination geometry. The molecule is extremely robust, allowing its deposition in a number of substrates at high temperatures conserving its molecular, electronic and magnetic characteristics.[79] Several reports demonstrate that this unit fulfils all requirements for the implementation spintronic quantum devices. In the following sections, we describe the characteristics which make $TbPc_2$ a viable nuclear spin molecular qubit with comparable and even superior parameters to alternative platforms.

*The $TbPc_2$ Molecular Qubit: A scalable well-defined system*

The SMM properties of $TbPc_2$ can be ascribed to the strong spin-orbit coupling of lanthanide ions and the ligand field exerted by the phthalocyanines, yielding a highly axial ground state characterised by an $|J = 6, J_z = \pm 6\rangle$ of the $^7F_6$ manifold.[81] Experimentally, it has been determined that the ground state, $|J = 6, J_z = \pm 6\rangle$, is separated from the first excited, $|J = 6, J_z = \pm 5\rangle$, state by ca. 600 K, with solely the ground doublet being populated at low temperatures (< 10 K) and low fields (< 10 T) (Fig. 4a).

The magnetic properties of $TbPc_2$ can be described through a Hamiltonian of the following form, $\mathcal{H} = \mathcal{H}_{lf} + g_J\mu_0\mu_B \boldsymbol{J} \cdot \boldsymbol{H} + A_{hf}\boldsymbol{I} \cdot \boldsymbol{J} + P(I_z^2 - \frac{1}{3}(I+1)\boldsymbol{I})$, where $\mathcal{H}_{lf}$ is the ligand field Hamiltonian ($\mathcal{H}_{lf} = \alpha B_2^0 O_2^0 + \beta(B_4^0 O_4^0 + B_4^4 O_4^4) + \gamma(B_6^0 O_6^0 + B_6^4 O_6^4)$) with $B_q^k$ representing the ligand field parameters, whilst α, β and γ are the Stevens constants. The second term represents the Zeeman energy, the third term accounts for hyperfine interactions and the fourth term is the quadrupole term. No quantum tunnelling of the magnetisation (QTM) is possible under $D_{4d}$. Experimentally, on the other hand, μ-SQUID measurements have shown several QTM events, which are ascribed to the lowering of the symmetry to $C_4$, which allows the presence of transverse anisotropy ($B_4^4 O_4^4 + B_6^4 O_6^4$) in the ligand field Hamiltonian. These terms induce mixing of the $|J_z = \pm 6\rangle$ states causing an avoided level crossing at zero field. This level crossing splits further in four due to strong hyperfine interaction between the $|J_z = \pm 6\rangle$ and $I = \frac{3}{2}$ of $Tb^{3+}$ causing four avoided level crossings (zoomed regions in Fig. 4b), leading to four new quantum numbers, i.e. $m_I = \pm \frac{3}{2}$ and $\pm \frac{1}{2}$. These have been proposed as nuclear spin qubits, due to their intrinsic isolated properties leading to long coherence times and low error rates.[34,35,82]

The tunnel splitting at these avoided level crossings are found to be $\Delta \approx 1$ μK as described by the Landau-Zener formula. Additionally, the presence of a quadrupole term $P$ causes an uneven separation between the $m_I$ states. It has been estimated that at zero external magnetic field a non-equal separation of $\nu_{01} \approx 2.45$ GHz, $\nu_{12} \approx 3.13$ GHz and $\nu_{23} \approx 3.81$ GHz separates the $m_I$ states. Likewise, the presence of a π-radical delocalised over the two Pc groups ($S = \frac{1}{2}$), which has been found to be ferromagnetically coupled to the $Tb^{3+}$ ion, is of utmost importance for the subsequent read-out. These features are of utmost importance for the independent manipulation of the nuclear spin states on TbPc$_2$ (*vide infra*).

*Readout*

QTM can occur at low field upon sweeping the magnetic field across the level crossings associated to $m_I = \pm^1/_2$ and $\pm^3/_2$. This causes a change in the electronic magnetic moment but preserves the nuclear spin. This feature has allowed the nuclear spin read-out through transport measurements having the TbPc$_2$ molecule suspended on carbon nanotubes (CNT)[32,33,83-85] and between gold junctions (Fig. 5).[34,35,82]

Read-out of the nuclear spin has been achieved employing an analogue of TbPc$_2$, where one of the Pc groups contains six hexyl groups and one 4-(4-pyren-1-ylbutoxy) group (TbPc$_2$*), allowing for improved grafting to the CNT. The molecules were suspended on CNTs in a spin valve configuration and the read-out was accomplished due to the strong interaction between the TbPc$_2$* unit(s) grafted to the wall of a CNT.[84] The detection was achieved through magneto-transport measurements. The results can be easily rationalised by considering, for simplicity, two TbPc$_2$* molecules suspended on the CNT. A maximum in the conductance is observed when the electronic spin of the TbPc$_2$* molecules are aligned in a parallel configuration (ferromagnetic coupling), whilst a minimum conductance is attained when the configuration is antiparallel (antiferromagnetic coupling). Transport experiments revealed distinct molecules lying on the CNT, each of them with different spatial orientation of the easy axes. Observation of the nuclear spin states for each TbPc$_2$* on the CNT was performed by measuring the tunnelling probability as a function of sweep rate, leading to the observation of four level crossings between ±50 mT, associated to the nuclear spin *I* = $^3/_2$ on Tb$^{3+}$. In these experiments, the read out of the nuclear spin was realised due to the QTM at the crossing levels, which allows for spin reversal of the SMM, taking place at low applied magnetic field ($\mu_0 H_z < \pm 50$ mT). Ganzhorn et al. demonstrated that QTM can be totally supressed at low temperatures for a single TbPc$_2$* suspended on a CNT due to one-dimensional phonons associated to the mechanical motion of the CNT. It has been proposed that this so-called quantum Einstein-de Hass effect could allow coherent spin manipulation.[83]

The first demonstration of electronic read-out of the nuclear states of Tb$^{3+}$, at a single molecule level, was accomplished after the TbPc$_2$ molecules was trapped into gold junctions, obtained by the electro-migration technique.[34] The read-out of the nuclear spins was achieved *via* indirect coupling where the current flows through a read-out dot (Pc in Fig. 4a and 5a), whilst the magnetic information is stored in the spin dot (the nuclear spins of the Tb$^{3+}$ ion). Due to the exchange coupled properties, the spin dot can influence the transport properties of the read-out dot, leading to the observation of the fingerprint magnetic properties of the Tb$^{3+}$ ion in the current passing through a single TbPc$_2$ molecule.[86]

Differential conductance studies (d*I*/d*V*) as a function of drain-source voltage ($V_{ds}$) and gate voltage ($V_g$) revealed a single charge-degeneracy point with a weak spin *S* = ½ Kondo effect, which is ascribed to the π-radical delocalised over the Pc rings. Since the *S* = ½ is ferromagnetically coupled to the magnetic moment carried by the Tb$^{3+}$ ion by ferromagnetic interaction, which is hyperfine coupled to the nuclear spin states, the transport properties through the aromatic Pc ligands (read-out dot) reflect the whole spin cascade $|S = 1/2\rangle||J = 6\rangle||I = 3/2\rangle$. In principle, this cascade effect allows sensing the nuclear spin. Consequently, Vincent et al. were able to read-out the single nuclear spin carried by the spin-dot employing experimental conditions close to a charge-degeneracy point, leading to a single abrupt jump in the differential conductance when sweeping the field from negative to positive values, which reversed when sweeping the field in the opposite direction (Fig 5b). If these conductance studies are continuously repeated, four different field positions for the conduction jumps are observed attributed to the reversal of the Tb$^{3+}$ nuclear spin magnetic moment, which slightly influences the read-out dot (Fig 5b,c). As observed, QTM at low field is highly efficient allowing the detection of the spin reversals at these four level crossings causing each time a change in the transport properties of the read-out dot.[34,35]

*Initialisation and Manipulation*

Upon alignment of the easy axis of TbPc$_2$ with the external magnetic field, ramping the magnetic field between ±60 mT, while monitoring the conductance jump of the readout dot, four QTM transitions are observed accounting for QTM events of the electronic spin occurring at avoided level crossings (Fig. 5b,c).

At low fields and temperatures, the initialisation of the four QTM events could be attained, corresponding to the nuclear qubit subspace of $|m_I = \pm^1/_2\rangle$ and $|m_I = \pm^3/_2\rangle$. The initialisation is achieved by sweeping the external magnetic field swept back and forth between ±60 mT until a QTM transition corresponding to the desired nuclear spin state is observed.[35,82] Thiele et al. reported the initialisation of the $|+^3/_2\rangle$ state and its controlled manipulation to a $|+^1/_2\rangle$ state, employing pulse sequences corresponding to the separation energy between these states. To this end, the authors made use of the hyperfine Stark effect, which is defined as the change of the hyperfine constant upon modulation of the external electrical field.[35] As example, we provide the reader with the description of the initialisation and manipulation sequence for the $|+^1/_2\rangle$ and $|+^3/_2\rangle$ states (Fig. 6a).

At low temperatures, solely the electronic ground state is populated. The strong spin orbit coupling then splits the $|J = 6\rangle$ state into four unevenly spaced microstates, owing to the quadrupolar interaction (Fig. 6a). To initialise the $|+^3/_2\rangle$ state, first the field is swept until a transition at −38 mT is observed, signalling the $|+^3/_2\rangle$ state. Once the transition has been observed at the specific magnetic field (in this case −38 mT), then the applied magnetic field is fixed (Fig. 6b). At this point, exploitation of the uneven spacing between the $\pm m_I$ states is accomplished through application of a radiofrequency pulse corresponding to the separation of the desired states. In the present case, the pulse has a frequency $\nu_{01} \approx 2.5$ GHz, with duration τ corresponding to the $|+^3/_2\rangle \leftrightarrow |+^1/_2\rangle$ subspace (Fig. 6c). The π pulse rotates $|+^3/_2\rangle$ to $|+^1/_2\rangle$. The final state is then detected by sweeping back the external magnetic field on a time scale faster than the measured relaxation times of both nuclear spin states, yielding a transition at −13 mT (Fig. 6d). Thiele and co-workers, finally observed Rabi oscillations and a nuclear qubit resonance frequency dependence on the gate voltage was found, attributed to the hyperfine Stark effect. The full initialisation-manipulation-read-out procedure has a duration of 2.4 s (Fig. 6,7).

As could be inferred, manipulation can additionally be achieved between the $|+^1/_2\rangle \leftrightarrow |-^1/_2\rangle$ states and $|-^1/_2\rangle \leftrightarrow |-^3/_2\rangle$ states by application of the appropriate resonance frequencies $\nu_{12}$ and $\nu_{23}$.

*Storage: spin-lattice relaxation and coherence times*

Another important aspect for qubits is the relaxation and coherence times, which must survive the required gate operation for a given calculation. In order to determine the relaxation times of the nuclear spin qubit, the real-time image of the nuclear spin trajectory was recorded after initialisation of the states as earlier described. Statistical analysis of the time at which each nuclear spin remained before changing to at different nuclear spin state allows the extraction of the spin-lattice relaxation times by fitting the data for an exponential form ($y = exp(-t/T_1)$) yielding relaxation times of $T_1 \approx 17$ s for $m_I = \pm^1/_2$ and $T_1 \approx 34$ s for $m_I = \pm^3/_2$ with fidelities of $F(m_I = \pm^3/_2) \approx exp(-5s/34s) \approx 93\%$ and $F(m_I = \pm^1/_2) \approx exp(-5/17s) \approx 87\%$ (Fig. 7a-d).[82]

Due to the facile initialisation, manipulation and read-out of the nuclear spin embodied in the TbPc$_2$ molecular qubit, the lifetime of the qubit, that is the duration of the coherence of the quantum superposition ($T_2^*$), was determined employing the Ramsey fringes. To this end, the nuclear spin qubit is firstly initialised at $|\pm m_I\rangle$ nuclear spin state as described in the initialisation procedure, followed by two $^\pi/_2$ microwave (MW) pulses with a interpulse delay $\tau$. The first $^\pi/_2$ pulse projects the $|\pm m_I\rangle$ spin into the x-y-plane. Precession of the spin around the z-axis at a given frequency ($\omega$) is obtained during the $\tau$ evolution time, while the second $^\pi/_2$ pulse brings the nuclear spin back into the x-z plane. Read-out is finally achieved as described as for the Rabi oscillation detection scheme resulting in Ramsey fringes (Fig. 9).

As observed in Fig. 9 the data follow an exponentially decaying cosine function yielding coherence time up to 0.32 ms. The determined values, although quite large, are smaller than expected can be expectedly improved.

*From Qubit to Qudit manipulation: Grover's Algorithm*

The realisation of Grover's algorithm employing SMM was proposed by Loss and Leuenberger[42] where the multilevel characteristics of SMMs would be exploited without requirement of inter-qubit interactions, that is entanglement. This algorithm is the succession of two gates. The first one, the Hadamard gate, starts from an initialised state to create a superposition of all qu*d*its states. Then, the Grover gate amplifies the amplitude of the researched state which has been previously labelled *via* its phase or its energy. As a result, making use of quantum amplitudes to determine the probabilities of an event, it is mostly probable to find the researched state. By operating on a highly superposed system, the Grover algorithm succeeds to quadratically speed up the amplitude amplification of the researched state, compared with a classical algorithm. Application of these quantum algorithms range from search in unsorted data bases, to pattern matching.[54]

In this sense, manipulation of each individual $m_I$ state contained in the TbPc$_2$ molecular qubit can be addressed *via* resonance frequencies leading to Rabi oscillations and during the determination of $T_2^*$ employing Ramsey fringes, both employing a single resonance frequency, therefore inducing a single $m_I$ transition. For the realisation of the Grover algorithm, manipulation of simultaneous $m_I$ states would be needed in order to create superposition of four nuclear spin states creating changing the qubit to a qu*d*it (with $d = 4$ for $I = ^3/_2$). To achieve this goal, a multifrequency pulse containing the resonance frequencies for each individual transition is obtained employing an arbitrary wave generator. Measurement of 3 and 4 nuclear spin states coherent superposition has been reported in a single TbPc$_2$ transistor.[87] Following this superposition, pulses parameters (frequencies and amplitudes) are tuned to reach a resonance condition in between the superposed states and the research state. As a result, the qu*d*it populations start to oscillate and the population of the labelled state briefly increase. The measurement of this population oscillation obtained after the Hadamard gate implementation is the first experimental proof of quantum algorithm implementation on a SMM (Fig. 10).[87]

## Conclusions

Single molecule magnets and their intriguing magnetic properties have led to their proposal in a variety of ambitious technological applications, promoting extensive investigation of these materials in assemblies and as discrete units. It could be shown that *single* TbPc$_2$ unit was embedded in scalable electronic circuits and *individual* spin read-out is performed by ligand-based read-out dots. Thereby, the spin-dot containing long-lived nuclear spin states is used to encode and process information at quantum level while the electronic magnetic state serves to address the quantum information and to transfer it to the read–out dot (and from there *via* the circuit to the external world). The TbPc$_2$-inherent spin cascade of $|S = 1/2\rangle||J = 6\rangle||I = 3/2\rangle$ decouples (and protects) the quantum information downwards, while it acts as an effective amplifier enabling the read-out of quantum information in upwards direction. These unique characteristics boosted a great deal of research leading to the observation of TbPc$_2$-SMMs fulfilling the requirements of the DiVincenzo criteria with very long coherence life times. In this sense, the TbPc$_2$ SMM, successfully perform the Grover algorithm where the manipulation of each individual $m_I$ state contained in the TbPc$_2$ molecular qubit can be addressed *via* resonance frequencies, inducing the desired $m_I$ transition. For the realisation of the Grover algorithm, simultaneous manipulation of $m_I$ states is achieved, allowing the creation superposition of the four nuclear spin states, embedded in the qu*d*it (with $d = 4$ for $I = ^3/_2$).

In consequence, we show that the molecular multilevel nuclear spin qubits meet practically all the essential characteristics to perform quantum operations, that is: (*i*) isolation, (*ii*) initialisation, (*iii*) read out, (*iv*) long coherence times and (*v*) manipulation, ultimately leading to the realisation of the Grover's algorithm. These results undoubtedly highlight the impressive characteristics of molecular materials.

Finally, although we have devoted this review entirely to the TbPc$_2$ molecule, where the π-radical delocalised over the Pc ligand plays a key role in the read out of the nuclear states, in molecules where the radical is absent, other read-out methods can be envisioned, such as by measuring the difference in cavity transmission[88], coupling the molecule to a photon emitter[89] or by transport measurements.[90-92] For such systems, the read-out schemes will entirely depend on the characteristics of the studied system.[93] Towards such goal molecular materials require certainly further studies. The rational design of molecular materials could ultimately allow the realisation of molecular devices working under quantum mechanical laws.

## Acknowledgements

We acknowledge the support of the Deutsche Forschungsgemeinschaft (DFG, TRR 88, "3MET", project C5) and FET-Open project "MOQUAS" by the European Commission, and the French Government through the ANR project "MolQuSpin".

# References

1. A. Einstein, B. Podolsky and N. Rosen, *Phys. Rev.*, 1935, **47**, 777–780.
2. E. Schrödinger, *Math. Proc. Camb. Phil. Soc.*, 1935, **31**, 555–563.
3. R. P. Feynman, *Int J Theor Phys*, 1982, **21**, 467–488.
4. D. Deutsch, *Proceed. Royal Soc. A*, 1985, **400**, 97–117.
5. P. W. Shor, *SIAM J. Comput.*, 1997, **26**, 1484–1509.
6. L. K. Grover, *Phys. Rev. Lett.*, 1997, **79**, 325–328.
7. S. Lloyd, *Science*, 1993, **261**, 1569–1571.
8. F. Jelezko, T. Gaebel, I. Popa, M. Domhan, A. Gruber and J. Wrachtrup, *Phys. Rev. Lett.*, 2004, **93**, 130501.
9. L. Childress, M. V. G. Dutt, J. M. Taylor, A. S. Zibrov, F. Jelezko, J. Wrachtrup, P. R. Hemmer and M. D. Lukin, *Science*, 2006, **314**, 281–285.
10. P. Neumann, J. Beck, M. Steiner, F. Rempp, H. Fedder, P. R. Hemmer, J. Wrachtrup and F. Jelezko, *Science*, 2010, **329**, 542–544.
11. D. Loss and D. P. DiVincenzo, *Phys. Rev. A*, 1998, **57**, 120–126.
12. K. C. Nowack, F. H. L. Koppens, Y. V. Nazarov and L. M. K. Vandersypen, *Science*, 2007, **318**, 1430–1433.
13. R. Prevedel, P. Walther, F. Tiefenbacher, P. Böhi, R. Kaltenbaek, T. Jennewein and A. Zeilinger, *Nature*, 2007, **445**, 65–69.
14. G. J. Milburn, *Phys. Scr.*, 2009, **T137**, 014003.
15. J. J. Pla, K. Y. Tan, J. P. Dehollain, W. H. Lim, J. J. L. Morton, F. A. Zwanenburg, D. N. Jamieson, A. S. Dzurak and A. Morello, *Nature*, 2013, **496**, 334–338.
16. J. J. Pla, K. Y. Tan, J. P. Dehollain, W. H. Lim, J. J. L. Morton, D. N. Jamieson, A. S. Dzurak and A. Morello, *Nature*, 2012, **489**, 541–545.
17. A. Laucht, R. Kalra, S. Simmons, J. P. Dehollain, J. T. Muhonen, F. A. Mohiyaddin, S. Freer, F. E. Hudson, K. M. Itoh, D. N. Jamieson, J. C. McCallum, A. S. Dzurak and A. Morello, *Nat. Nanotech.*, 2017, **12**, 61–66.
18. I. Chiorescu, Y. Nakamura, C. J. P. M. Harmans and J. E. Mooij, *Science*, 2003, **299**, 1869–1871.
19. J. Clarke and F. K. Wilhelm, *Nature*, 2008, **453**, 1031–1042.
20. I. Bloch, *Nature*, 2008, **453**, 1016–1022.
21. R. Blatt and D. Wineland, *Nature*, 2008, **453**, 1008–1015.
22. F. Troiani and M. Affronte, *Chem. Soc. Rev.*, 2011, **40**, 3119.
23. L. Bogani and W. Wernsdorfer, *Nat. Mater.*, 2008, **7**, 179–186.
24. J. M. Clemente-Juan, E. Coronado and A. Gaita-Ariño, *Chem. Soc. Rev.*, 2012, **41**, 7464.
25. G. Aromí, D. Aguilà, P. Gamez, F. Luis and O. Roubeau, *Chem. Soc. Rev.*, 2012, **41**, 537–546.
26. I. L. Chuang, L. M. K. Vandersypen, X. Zhou, D. W. Leung and S. Lloyd, *Nature*, 1998, **393**, 143–146.
27. L. M. K. Vandersypen, M. Steffen, G. Breyta, C. S. Yannoni, M. H. Sherwood and I. L. Chuang, *Nature*, 2001, **414**, 883–887.
28. M. Affronte, F. Troiani, A. Ghirri, A. Candini, M. Evangelisti, V. Corradini, S. Carretta, P. Santini, G. Amoretti, F. Tuna, G. A. Timco and R. E. P. Winpenny, *J. Phys. D: Appl. Phys.*, 2007, **40**, 2999–3004.
29. G. A. Timco, S. Carretta, F. Troiani, F. Tuna, R. J. Pritchard, C. A. Muryn, E. J. L. McInnes, A. Ghirri, A. Candini, P. Santini, G. Amoretti, M. Affronte and R. E. P. Winpenny, *Nat. Nanotech.*, 2009, **4**, 173–178.
30. D. Aguilà, L. A. Barrios, V. Velasco, O. Roubeau, A. Repollés, P. J. Alonso, J. Sesé, S. J. Teat, F. Luis and G. Aromí, *J. Am. Chem. Soc.*, 2014, **136**, 14215–14222.
31. J. Ferrando-Soria, E. Moreno Pineda, A. Chiesa, A. Fernandez, S. A. Magee, S. Carretta, P. Santini, I. J. Vitorica-Yrezabal, F. Tuna, G. A. Timco, E. J. L. McInnes and R. E. P. Winpenny, *Nat. Commun.*, 2016, **7**, 11377.
32. M. Urdampilleta, S. Klyatskaya, J.-P. Cleuziou, M. Ruben and W. Wernsdorfer, *Nat. Mate.*, 2011, **10**, 502–506.
33. M. Ganzhorn, S. Klyatskaya, M. Ruben and W. Wernsdorfer, *Nat. Nanotech.* 2013, **8**, 165–169.
34. R. Vincent, S. Klyatskaya, M. Ruben, W. Wernsdorfer and F. Balestro, *Nature*, 2012, **488**, 357–360.
35. S. Thiele, F. Balestro, R. Ballou, S. Klyatskaya, M. Ruben and W. Wernsdorfer, *Science*, 2014, **344**, 1135–1138.
36. J. N. Eckstein and J. Levy, *MRS Bull.*, 2013, **38**, 783–789.
37. M. Luo and X. Wang, *Sci. China Phys. Mech. Astron.*, 2014, **57**, 1712–1717.
38. M. Mohammadi, A. Niknafs and M. Eshghi, *Quantum Inf Process.*, 2010, **10**, 241–256.
39. D. P. O'Leary, G. K. Brennen and S. S. Bullock, *Phys. Rev. A*, 2006, **74**, 433.
40. D. C. P. Williams, in *Texts in Computer Science*, Springer London, London, 2011, pp. 51–122.
41. D. P. DiVincenzo, *Fortschritte der Physik*, 2000, **48**, 771–783.
42. M. N. Leuenberger and D. Loss, *Nature*, 2001, **410**, 789–793.
43. S. Lloyd, *Science*, 1996, **273**, 1073–1078.
44. P. W. Shor, *SIAM J. Comput.*, 1997, **26**, 1484–1509.
45. R. L. Rivest, A. Shamir and L. Adleman, *Commun. ACM*, 1978, **21**, 120–126.
46. T. Kleinjung, K. Aoki, J. Franke, A. K. Lenstra, E. Thomé, J. W. Bos, P. Gaudry, A. Kruppa, P. L. Montgomery, D. A. Osvik, H. t. Riele, A. Timofeev, and P. Zimmermann, in *Advances in Cryptology-CRYPTO 2010*, T. Rabin (ed), Springer, New York, USA, 2010, pp. 333.
47. A. G. Fowler, M. Mariantoni, J. M. Martinis and A. N. Cleland, *Phys. Rev. A*, 2012, **86**, 032324.
48. C.-Y. Lu, D. E. Browne, T. Yang and J.-W. Pan, *Phys. Rev. Lett.*, 2007, **99**, 250504.
49. E. Martín-López, A. Laing, T. Lawson, R. Alvarez, X.-Q. Zhou and J. L. O'Brien, *Nat. Photonics*, 2012, **6**, 773–776.
50. J. A. Jones, M. Mosca and R. H. Hansen, *Nature*, 1998, **393**, 344–346.
51. L. M. K. Vandersypen, M. Steffen, M. H. Sherwood, C. S. Yannoni, G. Breyta and I. L. Chuang, *Appl. Phys. Lett.*, 2000, **76**, 646.
52. T. van der Sar, Z. H. Wang, M. S. Blok, H. Bernien, T. H. Taminiau, D. M. Toyli, D. A. Lidar, D. D. Awschalom, R. Hanson and V. V. Dobrovitski, *Nature*, 2012, **484**, 82–86.
53. B. P. Lanyon, C. Hempel, D. Nigg, M. Müller, R. Gerritsma, F. Zähringer, P. Schindler, J. T. Barreiro, M. Rambach, G. Kirchmair, M. Hennrich, P. Zoller, R. Blatt and C. F. Roos, *Science*, 2011, **334**, 57–61.
54. A. Montanaro, *Quantum Inf.*, 2016, **2**, 15023.
55. M. J. Biercuk, H. Uys, A. P. VanDevender, N. Shiga, W. M. Itano and J. J. Bollinger, *Nature*, 2009, **458**, 996–1000.
56. C.-J. Yu, M. J. Graham, J. M. Zadrozny, J. Niklas, M. D. Krzyaniak, M. R. Wasielewski, O. G. Poluektov and D. E. Freedman, *J. Am. Chem. Soc.*, 2016, **138**, 14678–14685.
57. M. Shiddiq, D. Komijani, Y. Duan, A. Gaita-Ariño, E. Coronado and S. Hill, *Nature*, 2016, **531**, 348–351.
58. G. A. Timco, F. Tuna, E. J. L. McInnes, R. E. P. Winpenny, S. J. Blundell and A. Ardavan, *Phys. Rev. Lett.*, 2012, **108**, 107204.
59. K. Bader, D. Dengler, S. Lenz, B. Endeward, S.-D. Jiang, P. Neugebauer and J. V. Slageren, *Nat. Commun.*, 2014, **5**, 5304.
60. M. Atzori, L. Tesi, E. Morra, M. Chiesa, L. Sorace and R. Sessoli, *J. Am. Chem. Soc.*, 2016, **138**, 2154–2157.
61. A. Fedorov, L. Steffen, M. Baur, M. P. da Silva and A. Wallraff, *Nature*, 2011, **481**, 170–172.
62. L. M. Procopio, A. Moqanaki, M. Araújo, F. Costa, I. Alonso Calafell, E. G. Dowd, D. R. Hamel, L. A. Rozema, Č. Brukner and P. Walther, *Nat. Commun.*, 2015, **6**, 7913.
63. E. O. Kiktenko, A. K. Fedorov, A. A. Strakhov and V. I. Man'ko, *Physics Letters A*, 2015, **379**, 1409–1413.
64. O. Dp, G. Brennen and S. Bullock, *Phys. Rev. A,* 2006, *74,* 032334.62.
65. S. Balakrishnan, *Phys. Res. Int.*, 2014, **2014**, 1–5.
66. A. A. Popov, E. O. Kiktenko, A. K. Fedorov and V. I. Man'ko, *J. Russ. Laser Res.*, 2016, 37, 581–590.
67. E. O. Kiktenko, A. K. Fedorov, O. V. Man'ko and V. I. Man'ko, *Phys. Rev. A*, 2015, **91**, 042312.
68. P. Rungta, W. J. Munro, K. Nemoto, P. Deuar, G. J. Milburn and

C. M. Caves, in *Directions in Quantum Optics*, Springer Berlin Heidelberg, Berlin, Heidelberg, 2001, vol. 561, pp. 149–164.
69  F. Troiani, A. Ghirri, M. Affronte, S. Carretta, P. Santini, G. Amoretti, S. Piligkos, G. A. Timco and R. E. P. Winpenny, *Phys. Rev. Lett.*, 2005, **94**, 1–4.
70  A. Fernandez, E. Moreno-Pineda, C. A. Muryn, S. Sproules, F. Moro, G. A. Timco, E. J. L. McInnes and R. E. P. Winpenny, *Angew. Chem.*, 2015, **127**, 11008 –11011.
71  J. Ferrando-Soria, A. Fernandez, E. Moreno Pineda, S. A. Varey, R. W. Adams, I. J. Vitorica-Yrezabal, F. Tuna, G. A. Timco, C. A. Muryn and R. E. P. Winpenny, *J. Am. Chem. Soc.*, 2015, **137**, 7644–7647.
72  M. J. Graham, J. M. Zadrozny, M. Shiddiq, J. S. Anderson, M. S. Fataftah, S. Hill and D. E. Freedman, *J. Am. Chem. Soc.*, 2014, **136**, 7623–7626.
73  J. M. Zadrozny, J. Niklas, O. G. Poluektov and D. E. Freedman, *J. Am. Chem. Soc.*, 2014, **136**, 15841–15844.
74  M. S. Fataftah, J. M. Zadrozny, S. C. Coste, M. J. Graham, D. M. Rogers and D. E. Freedman, *J. Am. Chem. Soc.*, 2016, **138**, 1344–1348.
75  M. Atzori, E. Morra, L. Tesi, A. Albino, M. Chiesa, L. Sorace and R. Sessoli, *J. Am. Chem. Soc.*, 2016, **138**, 11234–11244.
76  L. Tesi, E. Lucaccini, I. Cimatti, M. Perfetti, M. Mannini, M. Atzori, E. Morra, M. Chiesa, A. Caneschi, L. Sorace and R. Sessoli, *Chem. Sci.*, 2016, **7**, 2074–2083.
77  M. J. Graham, J. M. Zadrozny, M. S. Fataftah and D. E. Freedman, *Chem. Mater.*, 2017, **29**, 1885–1897.
78  K. S. Pedersen, A.-M. Ariciu, S. McAdams, H. Weihe, J. Bendix, F. Tuna and S. Piligkos, *J. Am. Chem. Soc.*, 2016, **138**, 5801–5804.
79  E. Moreno Pineda, T. Komeda, K. Katoh, M. Yamashita and M. Ruben, *Dalton Trans.*, 2016.
80  P. C. Maurer, G. Kucsko, C. Latta, L. Jiang, N. Y. Yao, S. D. Bennett, F. Pastawski, D. Hunger, N. Chisholm, M. Markham, D. J. Twitchen, J. I. Cirac and M. D. Lukin, *Science*, 2012, **336**, 1283–1286.
81  N. Ishikawa, M. Sugita, T. Okubo, N. Tanaka, T. Iino and Y. Kaizu, *Inorg. Chem.*, 2003, **42**, 2440–2446.
82  S. Thiele, R. Vincent, M. Holzmann, S. Klyatskaya, M. Ruben, F. Balestro and W. Wernsdorfer, *Phys. Rev. Lett.*, 2013, **111**, 037203.
83  M. Ganzhorn, S. Klyatskaya, M. Ruben and W. Wernsdorfer, *Nat. Commun.*, 2016, **7**, 11443.
84  M. Urdampilleta, S. Klyatskaya, M. Ruben and W. Wernsdorfer, *Phys. Rev. B*, 2013, **87**, 195412.
85  M. Urdampilleta, S. Klayatskaya, M. Ruben and W. Wernsdorfer, *ACS Nano*, 2015, **9**, 4458–4464.
86  C. Godfrin, S. Thiele, A. Ferhat, S. Klyatskaya, M. Ruben, W. Wernsdorfer and F. Balestro, *ACS Nano*, 2017, **11**, 3984–3989.
87  C. Godfrin, A. Ferhat, R. Ballou, S. Klyatskaya, M. Ruben, W. Wernsdorfer and F. Balestro, 2017, **119**, 187702.
88  K. Hennessy, A. Badolato, M. Winger, D. Gerace, M. Atatuere, S. Gulde, S. Faelt, E. L. Hu and A. Imamoglu, *Nature*, 2007, **445**, 896–899.
89  W. E. Moerner and M. Orrit, *Science*, 1999, **283**, 1670–1676.
90  F. Schwarz, G. Kastlunger, F. Lissel, C. Egler-Lucas, S. N. Semenov, K. Venkatesan, H. Berke, R. Stadler and E. Lörtscher, *Nat. Nanotech.*, 2016, **11**, 170–176.
91  S. Wagner, F. Kisslinger, S. Ballmann, F. Schramm, R. Chandrasekar, T. Bodenstein, O. Fuhr, D. Secker, K. Fink, M. Ruben and H. B. Weber, *Nat. Nanotech.*, 2013, **8**, 1–5.
92  W. Y. Kim and K. S. Kim, *Acc. Chem. Res.*, 2010, **43**, 111–120.
93  S. Sanvito, *Chem. Soc. Rev.*, 2011, **40**, 3336.

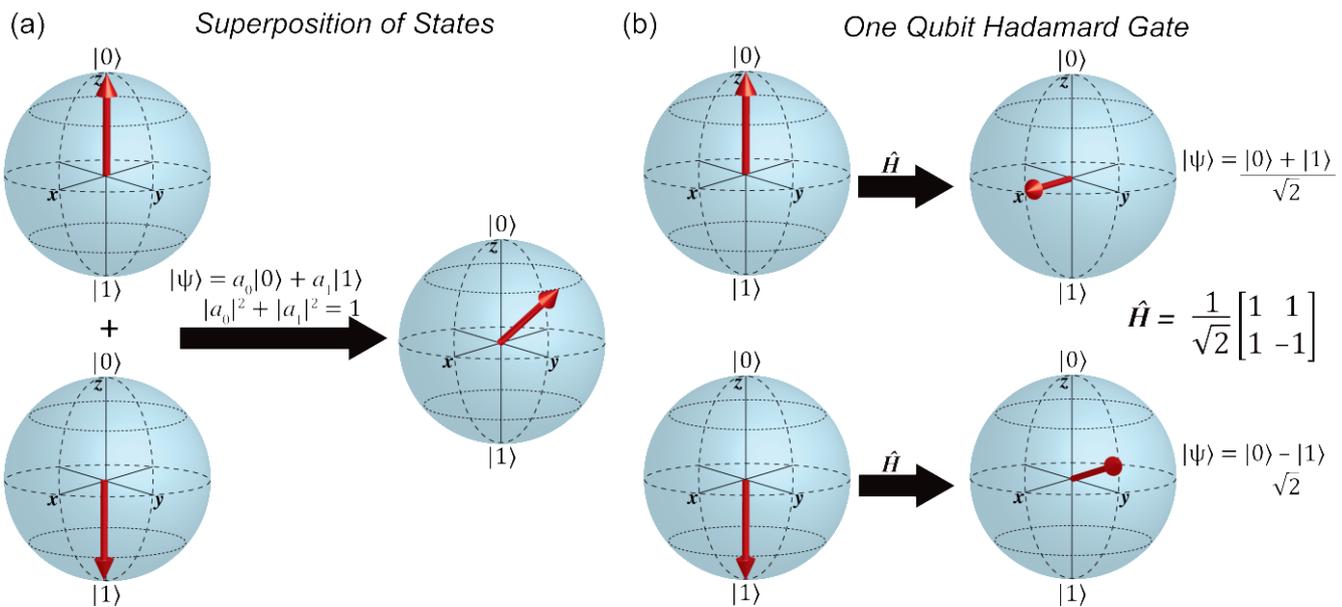

**Figure 1.** (a) The quantum version of the bit, a qubit, can be represented in the Bloch sphere with an arrow pointing north representing the $|0\rangle$ state, while when pointing south it represents the $|1\rangle$ state. Unlike the bit, the qubit can possess many more states, which can be viewed as an arrow pointing in any other direction of the sphere. These new states are quantum superposition of the $|1\rangle$ and $0\rangle$ states, giving the computational power expected in quantum computers; (b) One qubit Hadamard gate acting on an initial qubit. After each operation superposition of states are obtained, all of them containing all possible combinations of states.

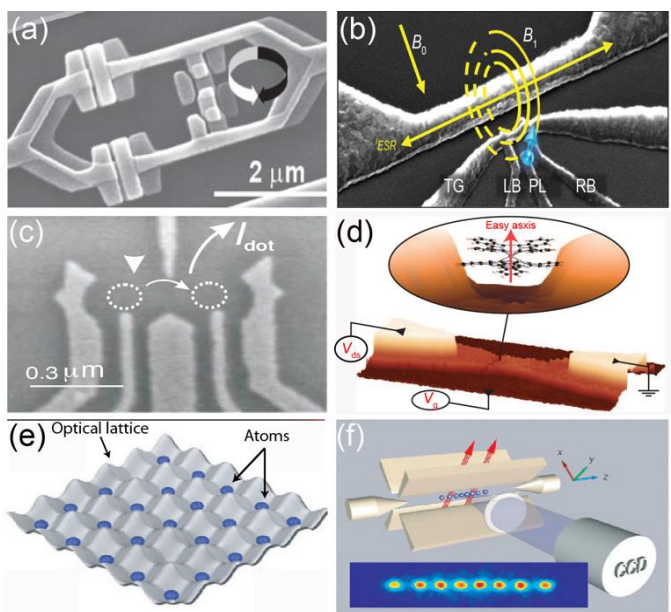

**Figure 2.** Quantum bits can be obtained from a wide range of material systems. Some examples of qubit materials are: (a) superconducting systems ("Reproduced from ref. 18 with permission from "The American Association for the Advancement of Science", copyright 2017); (b) impurities such as $^{31}P$ in Si ("Reproduced from ref. 16 with permission from "Nature Publishing Group", copyright 2017); (c) quantum dots ("Reproduced from ref. 12 with permission from "The American Association for the Advancement of Science", copyright 2017); (d) molecular qubits ("Reproduced from ref. 34 with permission from "Nature Publishing Group", copyright 2017); (e) ultracold atoms ("Reproduced from ref. 20 with permission from "Nature Publishing Group", copyright 2017) and (f) trapped ions ("Reproduced from ref. 21 with permission from "Nature Publishing Group", copyright 2017).

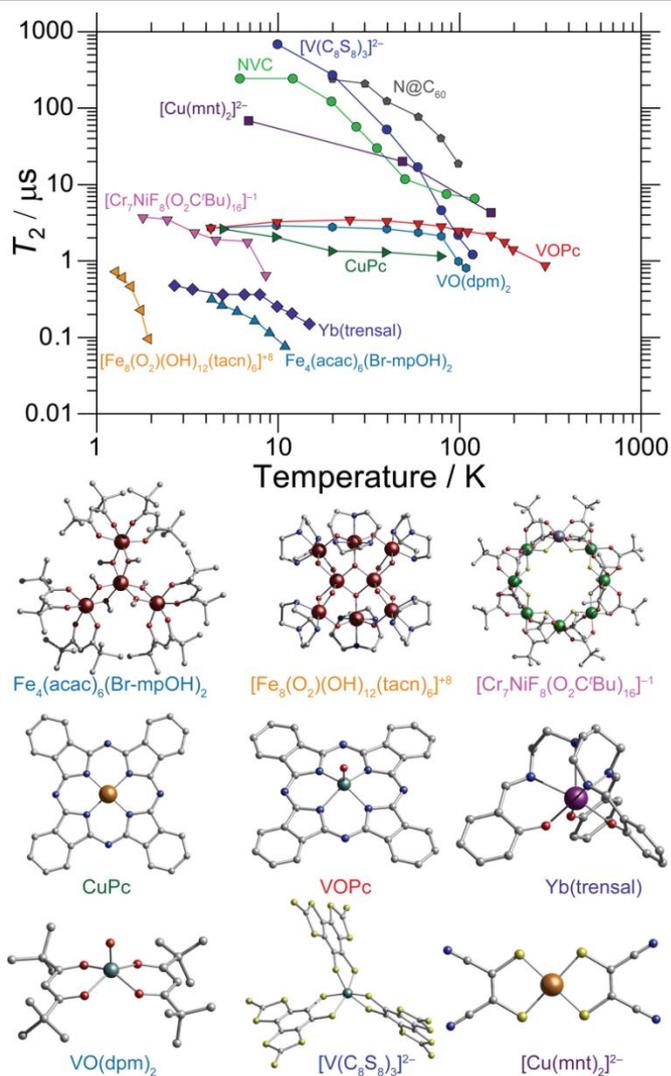

**Figure 3.** Electron spin-Spin relaxation($T_2^*$) versus temperature for several prospect qubits and the crystals structure of the complexes. Colour code: C, grey; S, yellow; N, blue; Fe, dark red; Cr, green; V, aqua; Cu, orange; Yb, purple; O, red; Ni, blue grey.

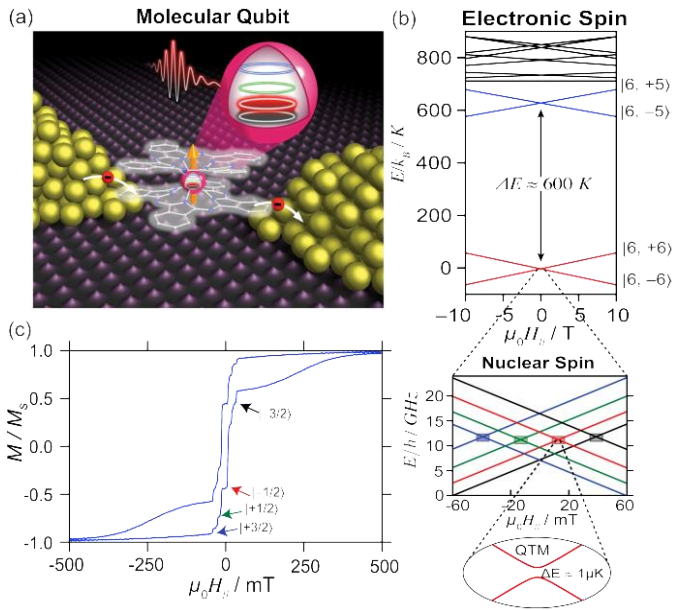

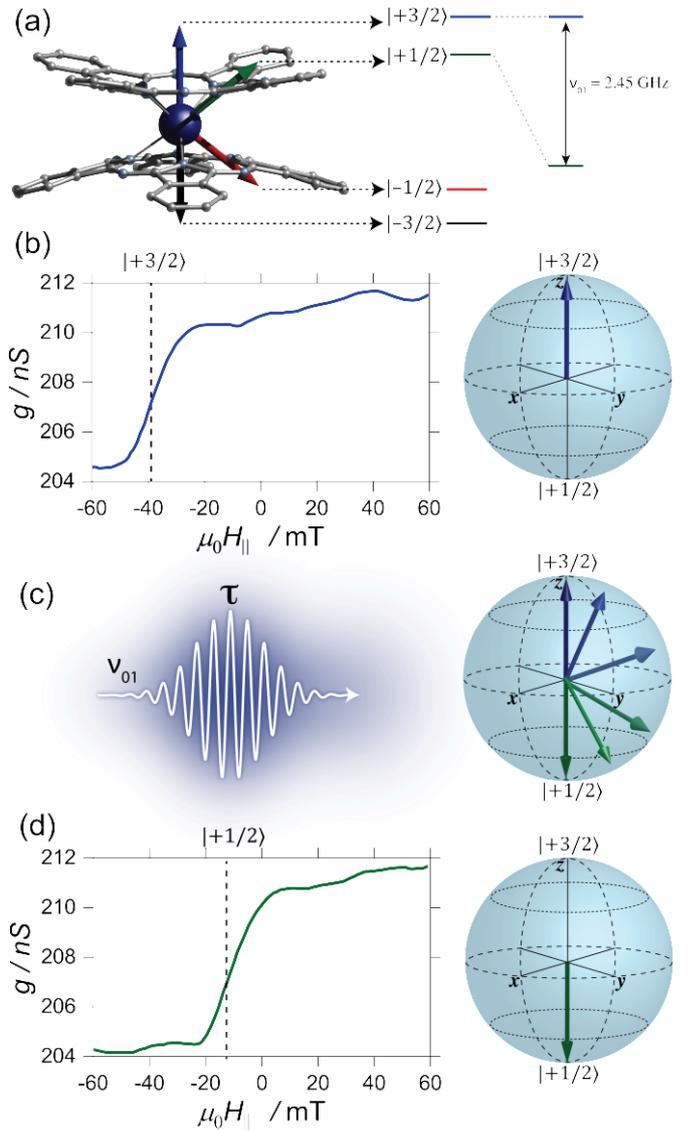

**Figure 4.** (a) Graphic representation of TbPc$_2$ deposited between two gold leads, while current flows through the molecule. (b) Energy level diagram resulting from strong spin orbit coupling of Tb$^{3+}$ and the ligand field exerted by the Pc groups. Zoomed regions shows the effect of strong hyperfine interaction which splits the $J_z = \pm 6$ state into four levels associated to $m_I = \pm^1/_2$ and $\pm^3/_2$ and avoided level crossing at due to mixing of states. (c) Hysteresis loop showing quantum tunnelling events associated to the nuclear spins. (Adapted from ref. 35. with permission from "The American Association for the Advancement of Science", copyright 2017)

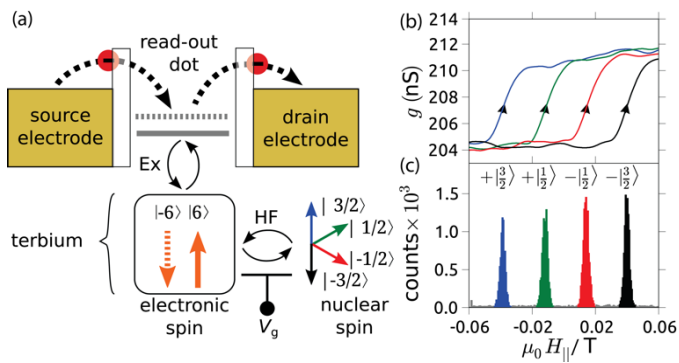

**Figure 5.** (a) schematic representation of a transistor with three coupled subsystems: (*i*) The four-level nuclear spin qubit is hyperfine (HF) coupled to (ii) an Ising-like electronic spin, which in turn is ferromagnetically exchange (Ex) coupled to (iii) a readout quantum dot. (b) Spin dependent conductance jumps of the readout quantum dot during magnetic-field sweeps. (c) Histograms of the positions of about 75,000 conductance jumps, showing four non-overlapping Gaussian-like distributions; each conductance jump can be assigned to a nuclear spin state. (Adapted from ref. 35. with permission from "The American Association for the Advancement of Science", copyright 2017)

**Figure 6.** (a) Bloch sphere representation of the nuclear spin of TbPc$_2$ and the energy separation between the nuclear spin states. (b-d) Graphical representation of initialisation, manipulation and detection of the $\left|+^3/_2\right\rangle \leftrightarrow \left|+^1/_2\right\rangle$ subspace. (b) (left) Field sweep conductance jump measurements with a transition corresponding to the $\left|+^3/_2\right\rangle$ state during the initialisation procedure at −38 mT and (right) Bloch sphere representation of the $\left|+^3/_2\right\rangle$ state. (c) (left) MW pulse with a duration τ and energy $\nu_{01} \approx$ 2.5 GHz and (right) precession of the $\left|+^3/_2\right\rangle$ state to $\left|+^1/_2\right\rangle$ state. (d) (left) Field sweep conductance jump measurement after the manipulation of the $\left|+^3/_2\right\rangle$ resulting in a transition at −13 mT, corresponding to the $\left|+^1/_2\right\rangle$ state and (right) Bloch sphere representation of the $\left|+^1/_2\right\rangle$ state after manipulation.

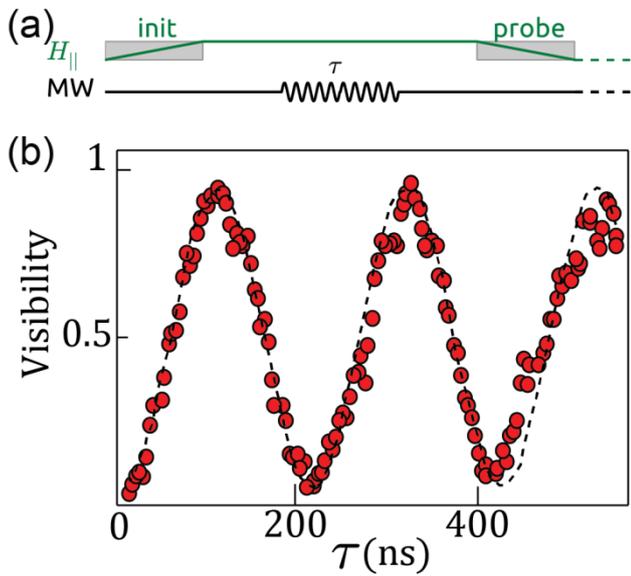

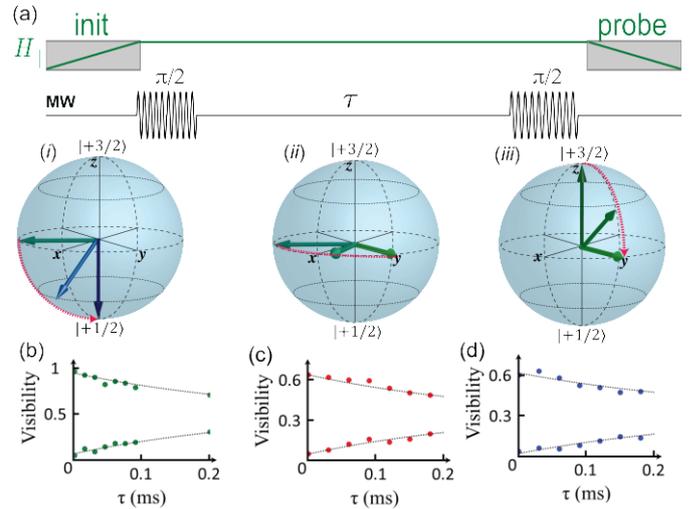

**Figure 7.** Nuclear spin manipulation. Rabi oscillation of the $|+3/2\rangle \leftrightarrow |+1/2\rangle$ subspace of the nuclear spin qubit. (a) Scheme of the initialisation-manipulation-probe sequence. First the nuclear spin $|+3/2\rangle$ is initialised, followed by a MW pulse of $\nu_{01}$ frequency (ca. 2.5 GHz) and duration $\tau$, inducing oscillating coherent manipulation of the two lower states of the nuclear spin qubit. As final step, the magnetic field is swept back to probe the final state. (b) Rabi oscillations between $|+3/2\rangle$ and $|+1/2\rangle$ states obtained by repeating the sequence 100 times for varying $\tau$ values, at two different MW powers.

**Figure 9.** Ramsey Fringes sequence: (a) Initialisation-manipulation-probe Ramsey fringes sequence. The following example involves the $|+3/2\rangle \leftrightarrow |+1/2\rangle$ subspace and can be applied to any other $|\pm m_I\rangle$ set (*i*) a pulse of $\nu_{01}$ frequency and duration $\tau = {}^{\pi}/_2$ is applied to a given nuclear spin $|+1/2\rangle$ projecting the spin into the equatorial plane. (*ii*) Precession of the spin into the x-plane during a $\tau$ time. (*iii*) a second ${}^{\pi}/_2$ pulse projects the y component of the spin state into the z-plane. The final state is finally determined *via* sweeping the field. (b-d) Experimental Ramsey fringes decay reveal coherence time values ($T_2^*$) of 0.28, 0.3 and 0.32 ms for the 1st, 2nd and 3rd, respectively. Adapted from ref. 87.

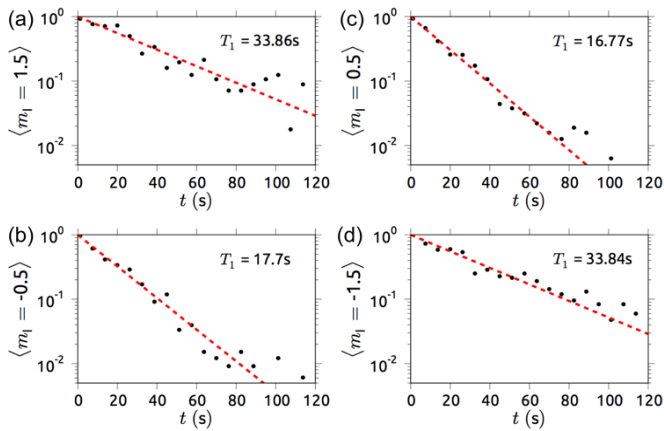

**Figure 8.** (a-e) Nuclear spin trajectory vs. time. Data fitted to an exponential decay yield $T_1$ for each independent nuclear spin, i.e. $T_1 = 17$ s for $|m_I = \pm 1/2\rangle$, $T_1 = 34$ s for $|m_I = \pm 3/2\rangle$. (Adapted from ref. 82. with permission from "American Physical Society", copyright 2017)

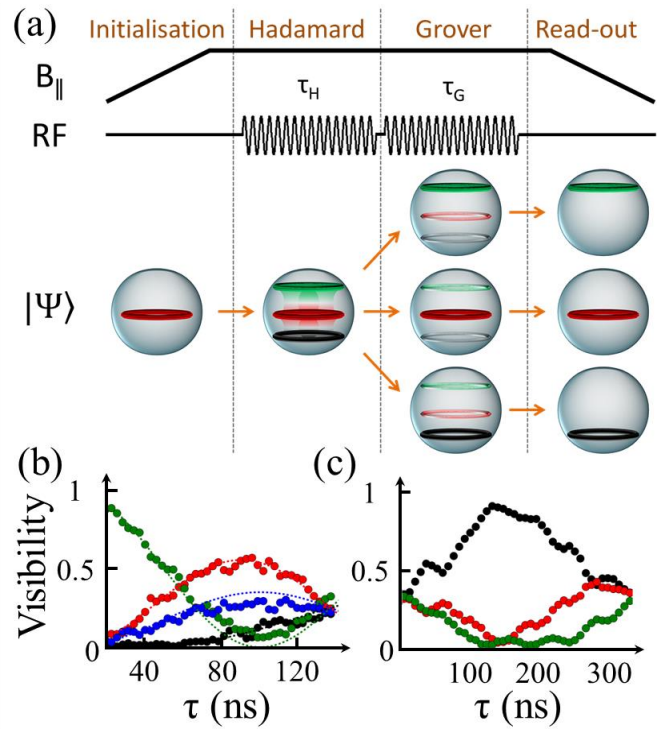

**Figure 10.** (a) the Grover algorithm is implemented using four different steps: initialization, Hadamard gate, Grover gate and final read-out. (b) is the evolution of the population in function of the Hadamard gate pulse length. Starting from the green state, after a pulse of duration 130ns, the population of all the states are equal. (c) is the evolution of the population in function of the Grover gate pulse length. Starting from a superposed state (obtain by an Hadamard pulse sequence) the system oscillated between this superposed state and a desired state (here the black one). This population oscillation is the fingerprint of the Grover algorithm implementation. Adapted from ref. 87.